\documentstyle[prl,aps,preprint]{revtex}

\tighten

\begin{document}
\draft
 
\pagestyle{empty}

\preprint{
\noindent
\hfill
\begin{minipage}[t]{3in}
\begin{flushright}
LBL--38110 \\
UCB--PTH--95/45 \\
SLAC--PUB--95--7089 \\
hep-ph/9601295 \\
January 1996
\end{flushright}
\end{minipage}
}

\title{Extracting $\bbox{R_b}$ and $\bbox{R_c}$ Without Flavor Tagging}

\author{
Jonathan L. Feng\thanks{Work supported in part by the Department of 
Energy under Contract DE--AC03--76SF00098 and in part by the National 
Science Foundation under grant PHY--90--21139.}
\thanks{Research Fellow, Miller Institute for Basic 
Research in Science.}
and Hitoshi Murayama\footnotemark[1]
}

\address{
Department of Physics,
University of California, Berkeley, California 94720\\
and\\
Theoretical Physics Group, Lawrence Berkeley Laboratory,
University of California, Berkeley, California 94720
}

\author{
James D. Wells\thanks{Work supported by the Department of Energy under Contract
DE--AC03--76SF00515.}
} 

\address{
Stanford Linear Accelerator Center,
Stanford University, Stanford, California 94309
}

\maketitle

\begin{abstract}
At present, two outstanding discrepancies between experiment and the
standard model are the measurements of the hadronic branching fractions
$R_b$ and $R_c$.  We note that an independent measurement of these
branching fractions may be obtained from the width of hadronic $Z$
decays with a prompt photon, $\Gamma_{q\bar{q}\gamma}$, along with the
total hadronic decay rate, $\Gamma_{\text{had}}$, and an additional
theoretical assumption.  Such an analysis requires no flavor tagging.
We consider several plausible theoretical assumptions and find that the
current value of $\Gamma_{q\bar{q}\gamma}$ favors larger $R_b$ and
smaller $R_c$ relative to standard model predictions, in accord with the
direct measurements.  If $\Gamma_{q\bar{q}\gamma}$ and
$\Gamma_{\text{had}}$ are combined with the direct measurements,
generation-blind corrections to all up-type and all down-type quark
widths are most favored.  An updated measurement of
$\Gamma_{q\bar{q}\gamma}$ with the currently available LEP data is
likely to provide an even stronger constraint on both the branching
fraction discrepancies and their possible non-standard model sources.
\end{abstract}
\pagestyle{plain}

\setcounter{footnote}{0}
\setlength{\baselineskip}{.23in}

The LEP and SLC $e^+e^-$ colliders have provided many impressive
confirmations of the standard model (SM) through high-precision studies
of the $Z$ boson.  At present, however, the combined average of direct
measurements of the $Z$ branching fractions $R_b \equiv
\Gamma_{b\bar{b}} / \Gamma_{\text{had}} $ and $R_c$ disagree 
with SM predictions at the level of $3.7\sigma$ and $2.3\sigma$,
respectively~\cite{EWWG,kniehl92:175}. These direct measurements rely
heavily on flavor tagging. It is therefore essential that the flavor
tagging efficiencies be calibrated accurately.  Impressive techniques
have recently been developed, including, most notably, the double-tag
method for calibrating $b$-tagging efficiency, which is limited
basically by statistics only.  However, given their status as two of the
most significant deviations from the SM, it is worth investigating
alternative methods for measuring $R_b$ and $R_c$ that are independent
of the systematic uncertainties inherent in the direct measurements.

A measurement of the decay width of prompt photon production in hadronic
$Z$ decays, which we denote $\Gamma_{q\bar{q}\gamma} \equiv \Gamma (Z\to
q\bar{q}\gamma)$, provides such an alternative.  The total hadronic
decay width is $\Gamma_{\text{had}} = \sum_{i=u,c,d,s,b} \Gamma_i$.  In
the width $\Gamma_{q\bar{q}\gamma}$, however, the up-type quark
contribution is enhanced, and so $\Gamma_{q\bar{q}\gamma} \propto 4
\sum_{i=u,c} \Gamma_i + \sum_{i=d,s,b}\Gamma_i$.  These two 
measurements, then, along with an assumption relating the light quark
widths to those of $b$ and $c$, provide flavor tagging
independent determinations of $R_b$ and $R_c$.  They may also provide
additional constraints on possible deviations from SM values.

By definition, events contributing to $\Gamma_{q\bar{q}\gamma}$ are
events in which the photon is radiated from a primary quark, {\it i.e.},
one of the two quarks that couples directly to the $Z$.  The
uncertainties in $\Gamma_{q\bar{q}\gamma}$ arise from backgrounds where
an isolated photon comes from other sources, {\it e.g.}, initial state
radiation and hadronization, and also from difficulties in the Monte
Carlo modeling~\cite{MC}.  A global average of results from currently
available analyses~\cite{LEPqqg} gives $R_{q\bar{q}\gamma} \equiv
\Gamma_{q\bar{q}\gamma}^{\text{SM}}/\Gamma_{q\bar{q}\gamma} = 1.077 \pm
0.042 \text{ (exp.)} \pm 0.04 \text{ (th.)}$~\cite{Mattig95}.  (Note that
$R_{q\bar{q}\gamma}$ is defined, following Ref.~\cite{Mattig95}, as the
theoretical value divided by the experimental value.)  It is interesting
to note that the current central value of $\Gamma_{q\bar{q}\gamma}$ is
about $1.3\sigma$ below the SM prediction.  

Given the currently available event sample of $\sim {\cal O}(10^7)$
hadronic $Z$ events, the statistical error may be reduced to $\sim
1\%$~\cite{Mattig93}.  The overall error would then be dominated by
systematic errors, which are primarily uncertainties in parton shower
modeling and $\alpha_s$ and have been estimated to be $\sim
3.5\%$~\cite{Mattig93}.  The total fractional error of
$\Gamma_{q\bar{q}\gamma}$ may therefore be improved from 5.8\% to $\sim
3.7\%$ after all the LEP data is analyzed~\cite{3percents}.  Such an
updated experimental analysis will increase the power of this study
considerably, as will be seen below.

To determine $R_b$ and $R_c$ from the two measurements
$\Gamma_{\text{had}}$ and $\Gamma_{q\bar{q}\gamma}$, it is clear that we
must choose a theoretically motivated framework for discussing
deviations from SM branching fractions.  We begin by parametrizing
possible shifts in the partial widths by the fractional deviations
$\delta_q$, defined by

\begin{equation}
\Gamma_q=\Gamma^{\text{SM}}_q \left( 1 + \delta_q \right) \ ,
\end{equation}
where $\Gamma_q$ is the partial width $\Gamma (Z\to q\bar{q})$, and
$\Gamma^{\text{SM}}_q$ is its SM value.  With this definition, the
shifts in the observables we will analyze are

\begin{eqnarray}
\delta R_i &= & \frac{\Gamma^{\text{SM}}_i \delta_i - R^{\text{SM}}_i
\sum_{q} \Gamma^{\text{SM}}_q \delta_q}
{\Gamma_{\text{had}}^{\text{SM}} + 
\sum_{q} \Gamma^{\text{SM}}_q \delta_q}\ ,\\
\delta \Gamma_{\text{had}} &= & 
\sum_{q} \Gamma^{\text{SM}}_q \delta_q  \ ,\\
\delta \Gamma_{q\bar{q}\gamma} &\propto & 
         4\sum_{f=u,c}{\Gamma^{\text{SM}}_f}\delta_f+
 	 \sum_{f=d,s,b}{\Gamma^{\text{SM}}_f}\delta_f \ ,
\label{Gqqgeqn}
\end{eqnarray}
where $q=u,c,d,s,b$, and $i=b,c$.  

The above parametrization accommodates a variety of new physics sources,
such as $Z$-$Z'$ mixing, new oblique corrections, and $Zq\bar{q}$ vertex
corrections.  Implicit in Eq.~(\ref{Gqqgeqn}), however, is the
assumption that the effects of new physics on the prompt photon width
are proportional to the primary quark charges, as is true when the
photon is radiated from a primary quark.  In general, this may be
violated, for example, by box diagrams in which the photon is attached
to an internal loop.  We assume, however, that the effects of such
diagrams are smaller than those of oblique and vertex corrections, as is
typically true in many new physics scenarios~\cite{box}.

At this stage, we have parametrized deviations from the SM in the five
parameters $\delta_q$.  To extract $R_b$ and $R_c$ from
$\Gamma_{\text{had}}$ and $\Gamma_{q\bar{q}\gamma}$, we must further
reduce the number of parameters to two.  We consider the following
scenarios, where the listed $\delta_q$ parameters are allowed to vary
subject to the given constraints, and {\it all unlisted $\delta_q$'s are
assumed to vanish}:
\par\quad
(I) $\delta_c$, $\delta_b$ ($c/b$ case) 
\par\quad
(II) $\delta_u = \delta_c$, $\delta_d = \delta_s = \delta_b$ 
(generation-blind case)
\par\quad
(III) $\delta_u = \delta_c$, $\delta_b$ ($uc/b$ case)
\par\quad
(IV) $\delta_b$ ($b$ case) \ .
\par

These scenarios are by no means exhaustive, but have a number of
interesting motivations.  The $c/b$ case is an obvious first choice, as
it is the most conservative scenario consistent with the anomalous
direct measurements of $R_c$ and $R_b$.  One should note, however, that
$\delta_u \not\simeq \delta_c$ and $\delta_d \not\simeq \delta_s$ are
each theoretically disfavored by the constraints from flavor-changing
neutral currents (FCNC).  Suppose the $Zc\bar{c}$ and $Zu\bar{u}$
couplings differ by $\epsilon\simeq 5\%$, as required to achieve a 10\%
reduction in $R_c$.  Suppose also that the mass eigenstates $u$ and $c$
are rotated by an angle $\theta$ relative to the interaction
eigenstates.  Let us consider the states $u_L$ and $c_L$. The rotation
generates the FCNC vertex $g_Z^u \epsilon \theta Z_\mu (\bar{u}_L
\gamma^\mu c_L) + \text{c.c.}$, where $g_Z^u \equiv e \left( 
\frac{1}{2} - \frac{2}{3} \sin^2 \theta_W\right)/ \sin\theta_W 
\cos\theta_W$, and $\theta_W$ is the weak mixing angle.  $Z$ boson 
exchange then generates a four-fermion operator $\frac{1}{2} \left(
g_Z^u \epsilon \theta / m_Z \right)^2 \bar{u}_L\gamma^\mu c_L
\bar{u}_L\gamma_\mu c_L$, which contributes to $D^0$--$\bar{D}^0$
mixing.  From the experimental bound $\Delta m_D < 1.3\times
10^{-13}$~GeV, one obtains a rough bound $\epsilon \theta \lesssim
3\times 10^{-4}$, or $\theta \lesssim 6\times 10^{-3}$ with $\epsilon =
0.05$, where we have taken $f_D^2 B_D \simeq (300$~MeV$)^2$.  A
difference in $\delta_d$ and $\delta_s$ is similarly constrained by
$K^0$--$\bar{K}^0$ mixing.  Simultaneous deviations from both $\delta_u
\simeq \delta_c$ and $\delta_d \simeq \delta_s$ are excluded.  These
arguments do not completely exclude the possibility of {\it either}\/
$\delta_u \not\simeq \delta_c$ {\it or}\/ $\delta_d \not\simeq
\delta_s$.  However, we see that, without some additional symmetries,
such possibilities require fine-tuning, and are therefore unnatural and
theoretically disfavored.

We are therefore led to consider scenarios with $\delta_u = \delta_c$
and $\delta_d = \delta_s$. The generation-blind case listed above is
perhaps the most well-motivated.  For example, a mixing between $Z$ and
a $Z'$ boson whose coupling is generation-blind leads to this case, as
do flavor-independent vertex corrections.  In addition, oblique
corrections depend only on quantum numbers, and so a scenario in which
oblique corrections are the dominant effect of new physics is an example
of the generation-blind case.  The $uc/b$ scenario is the most
conservative scenario that is consistent with both the LEP direct
measurements of $R_b$ and $R_c$ and the theoretical considerations of
the previous paragraph.

Finally, one can also consider the measured discrepancy in $R_c$ to be a
large statistical fluctuation and allow only $\delta_b$ to be
non-vanishing.  This scenario, the $b$ case, is realized if there is a
gauge boson that couples only to the third generation and mixes with the
$Z$ boson~\cite{holdom95:279},
or a large vertex correction to the $Zb\bar{b}$ vertex from
superparticles~\cite{Don} or technicolor~\cite{Chivukula}.  This
possibility could help resolve the longstanding difference between
$\alpha_s(m_Z^2)=0.123$ extracted from the $Z$ lineshape~\cite{EWWG} (in
the SM) and the lower $\alpha_s(m_Z^2)\simeq 0.110$ from many low energy
observables~\cite{shifman95:605}.  In fact, the change in $\alpha_s$ for
a given shift in the electroweak contribution to $\Gamma(Z\to b\bar b)$
is $\delta\alpha_s \simeq -0.7\delta_b$.  Using
$R_b^{\text{expt}}=0.2205\pm 0.0016$ when $R_c$ is fixed to its SM
value~\cite{EWWG}, a shift $\delta_b \simeq 0.02$ makes the measured and
SM predictions of $R_b$ consistent to about 1$\sigma$ and simultaneously
brings the value of $\alpha_s$ extracted from the $Z$ lineshape down to
about $0.110$.

For each of these scenarios, we now use the measured values of
$\Gamma_{\text{had}}$ and $\Gamma_{q\bar{q}\gamma}$ to determine $R_b$
and $R_c$, and we compare the extracted values of these branching
fractions to the direct measurements.  Table~\ref{table:1} shows the
measured values and SM predictions for $R_b$, $R_c$,
$\Gamma_{\text{had}}$, and $R_{q\bar q\gamma}$~\cite{kniehl92:175}.  In
applying the measured values of these quantities to constrain the
various scenarios, we assume that the new physics does not significantly
alter the detection efficiency of the prompt photon signal.  If it does,
the parameters $\delta_q$ and the efficiency are correlated, which
complicates the analysis.  However, as noted above, we assume that
oblique or vertex corrections are the dominant effects of new physics in
this analysis.  These corrections preserve all kinematical distributions
of the jets and photon for each quark chirality, and the efficiency is
therefore insensitive to the new physics effects.

The error for each of the observables is determined by adding in
quadrature the experimental measurement error and the uncertainties in
the top quark mass and strong coupling constant, which we take to be
$m_t=175\pm 15 \text{ GeV}$ and $\alpha_s(m_Z^2)=0.118\pm 0.006$.  Note
that we cannot use the value of $\alpha_s$ extracted from the global
fit, because we allow deviations of the widths $\Gamma_q$ from the SM.
The $\alpha_s$ measurements from low-energy data and jet shape variables
do not rely on electroweak physics, and so may be used in this analysis.

\begin{figure}
\psfig{file=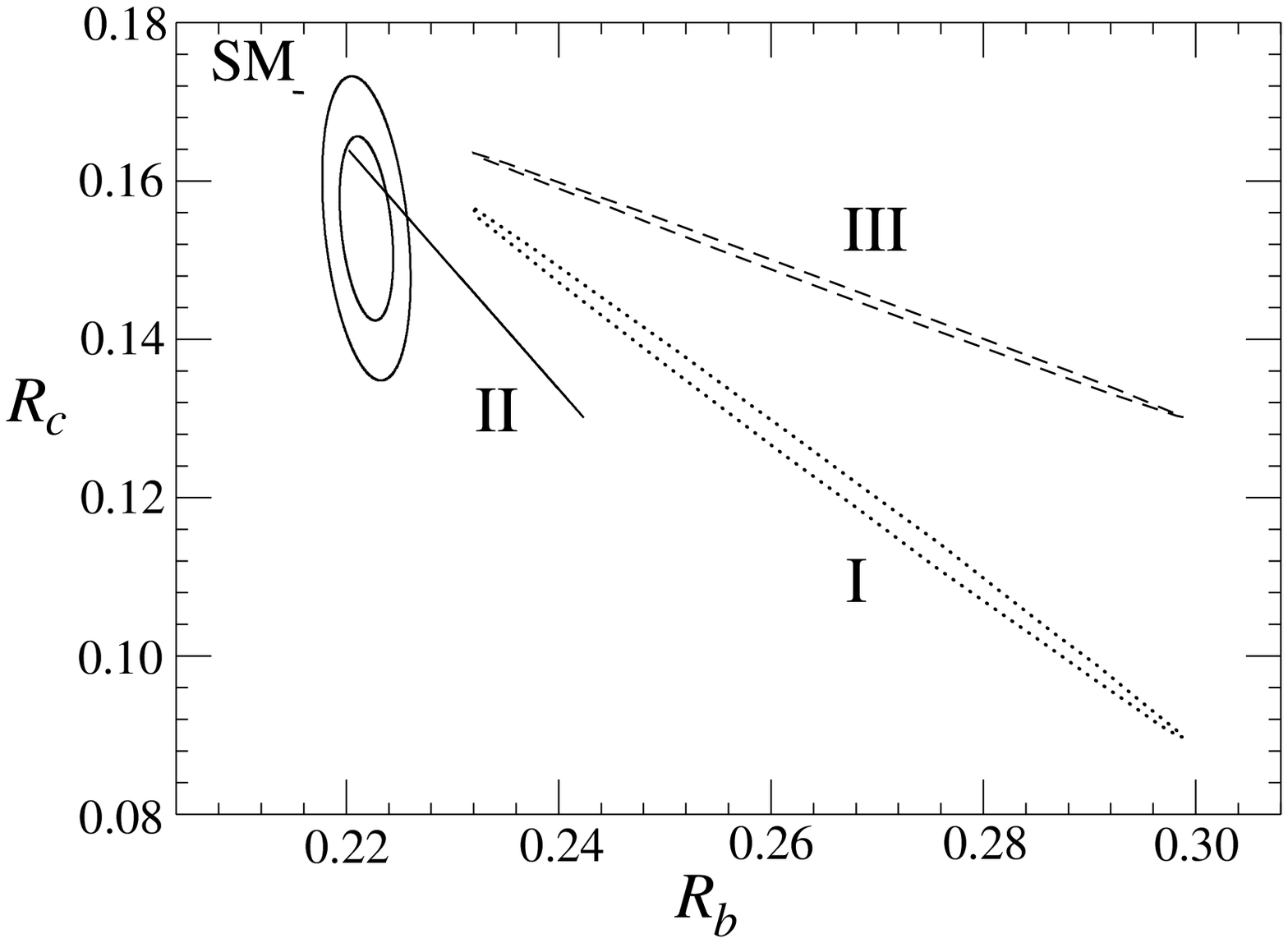,width=0.85\textwidth}
\caption{The $1\sigma$ allowed regions in the $(R_b, R_c)$ plane
extracted from the measured values of $\Gamma_{\text{had}}$ and
$\Gamma_{q\bar{q}\gamma}$ in the three scenarios: (I) $c/b$ case
(dotted), (II) generation-blind case (solid), and (III) $uc/b$ case
(dashed).  The ellipses are the 68\% and 95\% C.L. contours for the
direct measurements of $R_b$ and $R_c$, and the SM predictions, with
$m_t = 175 \pm 15$ GeV, are given by the very short line segment in the
upper-left corner.  The current value of $\Gamma_{q\bar{q}\gamma}$ has
an error of 5.8\%.  This is estimated to improve to 3.7\% given the
currently available LEP event samples, which will shrink the $1\sigma$
allowed regions by a factor of 0.64.
\label{fig:1}} 
\end{figure}

We present our results in Fig.~\ref{fig:1} for the extracted values of
$R_b$ and $R_c$ for each of the first three theoretical assumptions
discussed above.  (The $b$ case will be discussed below.)  For each
scenario, the measured values of $\Gamma_{\text{had}}$ and
$\Gamma_{q\bar{q}\gamma}$ determine a preferred region of the $(R_b,
R_c)$ plane.  The $1\sigma$ contours are plotted in Fig.~\ref{fig:1}.
All regions are long and narrow.  The width of each region is determined
by $\Gamma_{\text{had}}$, which is tightly constrained relative to the
other measurements, and the parametrization of the particular
theoretical scenario.  For example, in the generation-blind case, no
variation in the $\delta_q$ parameters changes $R_b$ without changing
$R_c$, so the associated band is very thin.  The slopes vary from case
to case because $\Gamma_{\text{had}}$ constrains different linear
combinations of $R_b$ and $R_c$ in the different scenarios.  The
positions of the regions are determined by the overlap of the
$\Gamma_{q\bar{q}\gamma}$ band with the $\Gamma_{\text{had}}$ band.  The
lengths are different for each case because the relative angle between
the two bands varies; if they are more parallel, the overlap region is
longer.

There are a number of interesting features of Fig.~\ref{fig:1}.  First
of all, it is noteworthy that the SM values for $R_b$ and $R_c$ are
outside the $1\sigma$ region for all scenarios.  This is a reflection of
the fact that the measured value of $\Gamma_{q\bar{q}\gamma}$ currently
differs from the SM prediction by $1.3\sigma$.  Second, for these
theoretical assumptions, the $1\sigma$ contours prefer higher $R_b$ and
lower $R_c$ than the SM values, because the measured
$\Gamma_{q\bar{q}\gamma}$ is below the SM prediction.  Since the error
in $\Gamma_{q\bar q\gamma}$ is much larger than that of
$\Gamma_{\text{had}}$, the lengths of the regions scale as the error in
$\Gamma_{q\bar q\gamma}$, and it is easy to see how the regions would
shrink as the accuracy in $\Gamma_{q\bar q\gamma}$ improves.  If the
error reduces to 3.7\% as expected given the currently available LEP
statistics discussed above, the lengths of the regions will decrease by
a factor of 0.64.  Depending on where the central value falls, the
measurement of $\Gamma_{q\bar{q}\gamma}$ may be quite significant.  For
example, if the central value were to remain at its present value, the
$\Gamma_{q\bar{q}\gamma}$ measurement would disagree with the SM at the
level of 2.1$\sigma$.

The $b$ case must be discussed separately since it has only one free
parameter.  What is interesting in this case is that one can extract
$R_b$ from $\Gamma_{\text{had}}$ alone, or from
$\Gamma_{q\bar{q}\gamma}$ alone.  These two extracted values can then be
compared to check the consistency of the scenario.  From
$\Gamma_{\text{had}}$ we obtain $R_b = 0.2170 \pm 0.0015$ ($0.2191 \pm
0.0015$) for $\alpha_s (m_Z) = 0.118$ ($0.110$).  On the other hand,
$\Gamma_{q\bar{q}\gamma}$ gives $R_b = 0.0877^{+0.1016}_{-0.0877}$.  The
extracted values of $R_b$ differ by about 1.3$\sigma$, irrespective of
the value assumed for $\alpha_s$.  A future improvement on
$\Gamma_{q\bar{q}\gamma}$ will certainly strengthen our ability to
determine this scenario's consistency.

We have also plotted in Fig.~\ref{fig:1} the 68\% and 95\% C.L. contours
for the direct measurements.  The combined measurement of all four
observables provides an opportunity to differentiate various new physics
scenarios. For example, it is evident from Fig.~\ref{fig:1} that the
direct measurements of $R_b$ and $R_c$ are most consistent with those
extracted from $\Gamma_{\text{had}}$ and $\Gamma_{q\bar{q}\gamma}$ in
the generation-blind case.  To quantify such a discussion, we now turn
to the results of global fits to all four observables for each of the
cases.  In the global fits, we treat the errors in $m_t$ and $\alpha_s$
as intrinsic uncertainties as before.  Alternatively, we could allow
$m_t$ and $\alpha_s$ to vary in the fits, but we choose to regard them as
uncertainties to simplify the discussion.  The correlation of $R_b$ and
$R_c$ in the direct measurements is also included.

For the the $c/b$, generation-blind, and $uc/b$ cases, we find that the
minimum $\chi^2/\mbox{d.o.f.}$ is 4.0/2, 1.6/2, and 5.1/2, respectively.
We find that the generation-blind case has no difficulty describing the
data, while the other cases are disfavored at more than 85\% C.L.
Indeed, it was shown that a mixing of $Z$ with an extra $E_6$ $U(1)$
gauge boson could improve the consistency between theory and
data~\cite{Hagiwara}.  Unfortunately, this particular realization of the
generation-blind case fails in the lepton sector, and such an
interpretation is excluded.  Nonetheless, our analysis clearly shows
that as a description of the hadronic widths and branching fractions,
the generation-blind case is the most favored of the new physics
scenarios we have considered.  For the $b$ case, there are two possible
attitudes.  If we fix $R_c$ at its SM predicted value and take
$\alpha_s=0.118$, a fit to $\Gamma_{\text{had}}$,
$\Gamma_{q\bar{q}\gamma}$ and $R_b^{\text{expt}}$ has a minimum
$\chi^2/\mbox{d.o.f.}$ of 4.0/2.  If we use the correlated experimental
values for both $R_c$ and $R_b$, the minimum is 8.3/3.  However if we
take $\alpha_s=0.110$, as advocated by Ref.~\cite{shifman95:605}, the
minimum $\chi^2/\mbox{d.o.f.}$ values for the two methods improve to
2.3/2 and 6.5/3, respectively.

Finally, we note that, imposing only the naturalness condition from the
FCNC considerations discussed above, the most general scenario allows
all $\delta_q$ parameters to vary subject to the constraints
\par\quad
(V) $\delta_u \simeq \delta_c$, $\delta_d \simeq \delta_s$, 
$\delta_b$ \ .
\par\noindent
This case is relevant if both large generation independent $\delta_q$
shifts, {\it e.g.}, shifts resulting from large non-standard oblique
corrections, and large $Z b\bar{b}$ specific corrections are present.
An analysis of such a case, however, is beyond the scope of this letter.

In conclusion, we find that the measurements of $\Gamma_{\text{had}}$
and $\Gamma_{q\bar{q}\gamma}$, when combined with a theoretical
assumption, provide a significant constraint on quark partial widths
without relying on flavor tagging.  In light of FCNC constraints, four
plausible theoretical assumptions were considered.  For each case, we
extracted $R_b$ and $R_c$ from $\Gamma_{\text{had}}$ and
$\Gamma_{q\bar{q}\gamma}$ and determined favored regions in the $(R_b,
R_c)$ plane.  The current measurement of $\Gamma_{q\bar{q}\gamma}$
prefers larger $R_b$ and smaller $R_c$ relative to SM predictions.
These regions, when compared with the direct determinations of $R_b$ and
$R_c$, may be used to help select among the many possible models of
physics beyond the SM. Of the four examples presented above, it appears
that generation-blind corrections provide a good fit to the data.
Scenarios in which only the $b$ and $c$ quark partial widths are allowed
to deviate from their standard model values are disfavored in this
analysis.  The analysis of all currently available LEP data is expected
to reduce the uncertainty in $\Gamma_{q\bar{q}\gamma}$ and will
significantly improve our ability to detect and interpret deviations
from the standard model.

\acknowledgements

We thank I.~Hinchliffe, M.~Mangano, P.~M\"attig, and Y.~Shadmi for
useful discussions and helpful correspondence.

\begin{table}
\caption{Measured and SM values (for $m_t=175\pm 15$ GeV) for four key 
observables.  $\Gamma_{\text{had}}$ is in GeV.  ``Pull'' is the
difference in the measured and SM central values in units of the
experimental error.}
\begin{tabular}{cccc} 
Observable & Measurement & Standard Model & Pull \\ \hline
$R_b$      & $0.2219\pm 0.0017$ & $0.2154\pm 0.0005$ & $+3.7$ \\
$R_c$      & $0.1540\pm 0.0074$ & $0.1711\pm 0.0002$ & $-2.3$ \\
$\Gamma_{\text{had}}$    
	   & $1.7448\pm 0.0030$ & $1.7405\pm 0.0039$ & $+1.4$ \\
$R_{q\bar q\gamma} $    & $1.077\pm 0.058$ & 1 & $+1.3$ 
\end{tabular}
\label{table:1}
\end{table}

\input psfig
\noindent
 

\begin{references}

\bibitem{EWWG}
The LEP Electroweak Working Group, LEPPEWWG--95--02 (1995).

\bibitem{kniehl92:175}
All SM partial widths are calculated with the aid of {\tt Z0POLE}.
B.~Kniehl, R.G.~Stuart, Comput. Phys. Comm. {\bf 72}, 175 (1992).

\bibitem{MC}
One potential difficulty lies in the modeling of the fragmentation of
quarks (and gluons) into photons, which is flavor-dependent.  Such
uncertainties may be greatly reduced by imposing isolation cuts to
remove most of the fragmentation region.  This approach and the errors
accompanying the introduction of isolation cuts are discussed in
Refs.~\cite{LEPqqg,Mattig95,Mattig93}.

\bibitem{LEPqqg}
ALEPH Collaboration, D. Buskulic {\it et al.}, Z. Phys. {\bf C57}, 17
(1993);
DELPHI Collaboration, P. Abreu {\it et al.}, CERN--PPE/95--101;
L3 Collaboration, O. Adriani {\it et al.}, Phys. Lett. {\bf B301}, 136
(1993); 
OPAL Collaboration, P. D. Acton {\it et al.}, Z. Phys. {\bf C58}, 405
(1993).

\bibitem{Mattig95}
P. M\"attig, invited talk at the International Symposium on Vector Boson
Self Interactions, Los Angeles, February 1--3, 1995, CERN--PPE/95--081.

\bibitem{Mattig93}
P. M\"attig, H. Spiesberger, and W. Zeuner, Z. Phys. {\bf C60}, 613
(1993).

\bibitem{3percents}
It is also possible that the high statistics of the currently available
event sample may be used to reduce the systematic errors.  However, it
is unlikely that the measurement may be improved beyond the 3\% level,
as measurements of the rate for three-jet events (or, in other words,
``prompt gluon production'') suffer from systematic uncertainties in
hadronization at this level.  See, for example, K.~Abe {\it et al.},
Phys. Rev. Lett. {\bf 71}, 2528 (1993).

\bibitem{box}
Diagrams that violate Eq.~(\ref{Gqqgeqn}) are typically suppressed by
the effective new physics scale $\Lambda$.  For example, box diagrams
lead to effective contact interactions like $e F^{\mu\nu} \bar{q}
\gamma_\mu q Z_\nu / \Lambda^2$.  In contrast, one may have
non-decoupling effects in oblique and vertex corrections.  Furthermore,
oblique and vertex corrections are enhanced both by internal quark
propagators, even after isolation cuts, and by the fact that they are in
phase with the corresponding tree-level amplitudes; such enhancements are
missing for the other diagrams.

\bibitem{holdom95:279}
See, for example, B.~Holdom, Phys. Lett. {\bf B351}, 279 (1995).  If the
mixing shifts $m_Z$ substantially, it modifies the overall normalization
of the decay rates $G_F m_Z^3$ and the weak mixing angle, resulting in
apparent shifts $\delta_u = \delta_c$, $\delta_d = \delta_s$, and
$\delta_b$.  This is the most general case naturally compatible with
FCNC constraints, case (V).

\bibitem{Don}
A. Djouadi {\it et al.}, Nucl. Phys. {\bf B349}, 48 (1991); M. Boulware
and D. Finnell, Phys. Rev. {\bf D44}, 2054 (1991).

\bibitem{Chivukula}
R. S. Chivukula, S. B. Selipsky, and E. H. Simmons,
Phys. Rev. Lett. {\bf 69}, 575 (1992).

\bibitem{shifman95:605}
M.~Shifman, Mod. Phys. Lett. {\bf A10}, 605 (1995); hep-ph/9511469.

\bibitem{Hagiwara}
K.~Hagiwara, plenary talk presented at the International Symposium on
Lepton Photon Interactions, Beijing, August 10--15, 1995,
hep-ph/9512425.

\end{references}
\end{document}